\documentclass[10pt]{article}
\usepackage{graphics}

\newcommand{\ket}[1]{|#1\rangle}
\newcommand{\inner}[2]{(\vec{#1},\vec{#2})}

\title{\textbf{Quantum Oracle Interrogation:}\\
Getting All Information for Almost Half the Price}
\author{\textit{Wim van Dam}\\
Centre for Quantum Computation, 
University of Oxford\\
Clarendon Laboratory, Parks Road, 
Oxford~~OX1~3PU,
U.K.\thanks{
\copyright 1998 IEEE. 
Published in the Proceedings of FOCS'98, 8-11 November 1998 
in Palo Alto, CA. 
Personal use of this material is permitted. 
However, permission to reprint/republish this material for advertising or 
promotional purposes or for creating new collective works for resale or 
redistribution to servers or lists, or to reuse any copyrighted component 
of this work in other works, must be obtained from the \textsc{ieee}. 
Contact: Manager, Copyrights and Permissions/\textsc{ieee} Service Center/ 
445 Hoes Lane/P.O.\,Box~1331 / Piscataway, NJ~08855-1331, USA.
Telephone:~+Intl.~732-562-3966.}
\and 
Quantum Computing and Advanced Systems Research, C.W.I.\\
P.O.\,Box 94079, NL--1090 GB, Amsterdam, The Netherlands\\
wim.van.dam@qubit.org}
\begin{document}
\maketitle
\begin{abstract}
Consider a quantum computer in combination with a binary oracle of
 domain size $N$. It is shown how $N/2+\sqrt{N}$ calls to the 
oracle are sufficient to guess the whole content of the oracle 
(being an $N$ bit string) with probability greater than 95\%. 
This contrasts the power of classical computers which would require
$N$ calls to achieve the same task.
From this result it follows that any function with the $N$ bits of 
the oracle as input can be calculated using $N/2+\sqrt{N}$ 
queries if we allow a small probability of error. It is also shown that 
this error probability can be made arbitrary small by using 
$N/2+O(\sqrt{N})$ oracle queries.  

In the second part of the article `approximate interrogation' is
considered. This is when only a certain fraction of the $N$ oracle 
bits are requested. Also for this scenario does the quantum algorithm
outperform the classical protocols. An example is given where
a quantum procedure with $N/10$ queries returns a string of which
80\% of the bits are correct. Any classical protocol would need
$6N/10$ queries to establish such a correctness ratio.  
\end{abstract}

\section{Introduction}
Recent research \cite{Beals,Farhi,Nayak} in quantum computation complexity 
has revealed several lower bounds on the capability of quantum computers to 
outperform classical computers in the black-box setting.
These results were proven by investigating the required amount of
queries to a black-box or oracle (with a domain size $N$) in order 
to decide some 
general property of this function. For example, if we want to know the 
parity of the $N$ black-box values with bounded error then it is 
still necessary for a quantum computer to call the black-box $N/2$ 
times\cite{Beals,Farhi}. It has also been shown that for the
\emph{exact\/} calculation of certain functions (the bitwise \textsc{or} 
for example) all $N$ calls are required\cite{Beals}.

This paper on the other hand, 
presents an \emph{upper bound\/} on the number of 
black-box queries that is necessary to compute any function
over the $N$ bits if we allow a small probability of error.
More specifically, 
it will be shown that for every unknown oracle
there is a potential speed-up of almost a factor of two if we 
want to know everything there is to know about the oracle function.
By this the following is meant.
If the domain of the oracle has size $N$, a classical computer will
have to apply $N$ calls in order to know all $N$ bits describing the 
oracle.
Here, it will be proven that a quantum computer can perform the same task 
with high probability using only $N/2+\sqrt{N}$ oracle calls.
From this result it follows immediately that \emph{any\/} 
(not necessarily binary) function $F$ on the domain
${\{0,1\}}^N$ can be calculated with a small 
two-sided error using only $N/2+\sqrt{N}$ calls. 

The factor--of--two gain can be increased by going to approximating
interrogation procedures. If we do not longer require to know
\emph{all\/} of the $N$ bits but are instead already satisfied with 
a certain percentage of correct bits, then the difference between
classical and quantum computation becomes bigger than for the 
above `exact interrogation' case. An example of this is when we 
want to guess the string such that we can expect 80\%  
of the bits to be correct. A quantum computer can do this 
with one-sixth of the queries that a classical computer requires
($N/10$ versus $3N/5$ calls). This also illustrates that the
procedure described here is not a `superdense coding--in--disguise'
which would only allow a reduction by a factor of two\cite{Bennett}. 

\section{Preliminaries}
The setting for this article is as follows. We try to investigate the 
potential differences between a quantum and a classical computer 
when both cases are confronted with an oracle $\omega$. The only thing 
known in advance about this $\omega$ is that it is a binary-valued 
function with a domain of size $N$. The oracle can therefore
be described by an $N$-bit string: 
$\vec{\omega} = \omega_1 \omega_2 \cdots \omega_N \in \{0,1\}^N$.   
The goal for both computers is to obtain the whole string $\vec{\omega}$ 
with high probability with as few oracle calls to $\omega$ as possible. 
The phrase ``with high probability'' means that the final answer of
the algorithm should be \emph{exactly\/} $\vec{\omega}$ 
at least $95\%$ of a time, 
for any possible $\omega$. Note that we are primarily concerned with the 
complexity of the algorithm in terms of oracle calls, both the time and 
space requirements of the algorithms are not considered when analysing the 
complexity differences. The model of an oracle as it used here goes also 
under the name of black-box, or database-query model.

The suggested quantum algorithm uses two procedures which are well-known
in quantum algorithm theory.
For reasons of clarity those two procedures will be explained in this
section before the actual algorithm is described.
We assume that the reader is familiar with the basics of quantum 
computation\cite{Berthiaume,Cleve}.

\subsection{One-Call Phase Kickback Trick}
Take an unknown function value $f\in\{0,1\}$. We want to induce
a phase $(-1)^f$ while calling the function only once. This can be
done in the following way (as described by Cleve \emph{et al.\/}\cite{Cleve}). 

If we start in the state $\ket{\Psi}(\ket{0}-\ket{1})/\sqrt{2}$
and we add (modulo 2) the value of $f$ to the last bit, then
the outcome will be 
\begin{eqnarray} \label{eq:pkb}
\ket{\Psi}\frac{\ket{0\oplus f}-\ket{1\oplus f}}{\sqrt{2}} & = & 
\left\{{\begin{array}{ll} 
+\ket{\Psi}\frac{\ket{0}-\ket{1}}{\sqrt{2}} & 
\textrm{if $f=0$,}\\
& \\
-\ket{\Psi}\frac{\ket{0}-\ket{1}}{\sqrt{2}} & 
\textrm{if $f=1$.}
\end{array}}\right.
\end{eqnarray} 
This correctly induces the phase $(-1)^f$ to the initial state.

\subsection{Inner Product versus Hadamard Transform} \label{sec:innerH}
The \emph{Hadamard transform $H$\/} is the one-qubit rotation that
maps $\ket{0}$ to the state $H\ket{0}=(\ket{0}+\ket{1})/\sqrt{2}$,
and the state $\ket{1}$ to $H\ket{1}=(\ket{0}-\ket{1})/\sqrt{2}$.
By the \emph{inner product\/} between two bit strings $\vec{x}$ and
$\vec{y}$, we mean the inner product modulo 2, that is:
\begin{eqnarray}
\inner{x}{y} & = & 
(x_1\cdot y_1)\oplus\cdots\oplus(x_N\cdot y_N)~.
\end{eqnarray}
This value can also be viewed as the parity of a subset of the
bit string $y_1 \cdots y_N$. This subset is described by the
characteristic vector $\vec{x}$ and its size equals the Hamming weight
$\|\vec{x}\|$ of the bit string $x_1\cdots x_N$. 
 
The Hadamard transform of a sequence of bits $y_1\cdots y_N$ and
the inner product function are closely related to each other:
for any $\vec{y} \in \{0,1\}^N$ it holds that
\begin{eqnarray} \label{eq:Hinner}
H^{\otimes N}\ket{\vec{y}} & = & 
\frac{1}{\sqrt{2^N}}
\sum_{\vec{x} \in \{0,1\}^N}{{(-1)}^{\inner{x}{y}}\ket{\vec{x}}}~.
\end{eqnarray}
Because $H$ is its own inverse, we can apply again a sequence of $N$ 
Hadamard transforms on the state in Equation~\ref{eq:Hinner} and thus
 obtain the original bit string $y_1\cdots y_N$ again:
\begin{eqnarray} \label{eq:Hinv}
H^{\otimes N}\left({
\frac{1}{\sqrt{2^N}}
\sum_{\vec{x} \in \{0,1\}^N}{{(-1)}^{\inner{x}{y}}\ket{\vec{x}}}
}\right) & = & 
\ket{\vec{y}}~.
\end{eqnarray}
The above leads to the observation that if we want to know the string 
$y_1\cdots y_N$, it is sufficient to have a superposition with phase 
values of the form $(-1)^{\inner{x}{y}}$, for every $\vec{x} \in \{0,1\}^N$. 
This is a well-known result in quantum computation and has been used several
times \cite{Bernstein,Cleve,Grover,Terhal} to underline the differences between
quantum and classical information processing.

\section{The Quantum Algorithm} \label{section:quantum_algorithm}
\subsection{Outline of the Algorithm}
The algorithm presented here is an approximation of the procedure described
in the Equations~\ref{eq:Hinner} and \ref{eq:Hinv}. Instead of calculating
the phase values $(-1)^{\inner{x}{\omega}}$ for \emph{all\/} 
$\vec{x} \in \{0,1\}^N$, we will do this only for the strings 
$x_1 \cdots x_N$ which do not have a Hamming weight (the number of ones 
in a bit string) above a certain threshold $k$.
By doing so, we can reduce the number of necessary oracle calls while
obtaining an outcome which is still very close to the `perfect state' as shown
in Equation~\ref{eq:Hinner} (but now for $\vec{\omega}$ instead of $\vec{y}$).

As stated in Section \ref{sec:innerH}, the value $\inner{x}{\omega}$
corresponds to the parity of a set of $\omega_i$ bits, where this set is
determined by the ones in the string $x_1 \cdots x_N$. 
To calculate the parity we can perform a sequence of additions
modulo 2 of the relevant $\omega_i$ values, 
where each $\omega_i$ has to be (and can be) obtained by one oracle call. 
It is therefore that the Hamming weight $\|\vec{x}\|$ equals the 
`oracle call complexity' of the reversible procedure 
(for an arbitrary bit $b \in \{0,1\}$):
\begin{eqnarray}
\ket{\vec{x}}\ket{b} & 
\stackrel{\|\vec{x}\|\mathrm{~oracle~calls}}{\longrightarrow} &
\ket{\vec{x}}\ket{b \oplus \inner{x}{\omega}}~.
\end{eqnarray}
The number of oracle calls will be reduced by the usage of a threshold
number $k$, such that we only compute the parity value $\inner{x}{\omega}$ 
if the Hamming weight of $\vec{x}$ is less than or equal to $k$. 
The algorithm
that performs this conditional parity calculation is denoted by
$A_k$ and its behaviour is thus defined by:
\begin{eqnarray} \label{eq:Ak}
A_k\ket{\vec{x}}\ket{b} & = & 
\left\{{
\begin{array}{ll}
\ket{\vec{x}}\ket{b \oplus \inner{x}{\omega}} & 
\textrm{ if $\|\vec{x}\|\leq k$,}\\
\ket{\vec{x}}\ket{b} & 
\textrm{ if $\|\vec{x}\|>k$.}
\end{array}
}\right. 
\end{eqnarray}
It is important that this algorithm $A_k$ requires at most $k$ oracle calls 
for every $x_1\cdots x_N$.
Because $A$ is reversible and does not induce any undesired phase changes,
we can also apply it to a superposition of different $\vec{x}$ strings. 
This brings us finally to the actual algorithm.

\subsection{The Actual Algorithm}
Prepare the state $\Psi_k$ which is an equally weighted superposition of 
bit strings of size $N$ with Hamming weight $\|\vec{x}\|$ less than or 
equal to $k$ and an additional qubit in the
state  $(\ket{0}-\ket{1})/\sqrt{2}$ attached to it:
\begin{eqnarray} \label{eq:Psi}
\ket{\Psi_k}\frac{\ket{0}-\ket{1}}{\sqrt{2}} & = &
\frac{1}{\sqrt{M_k}}
\left\{{\sum_{\vec{x}\in \{0,1\}^N}^{\|\vec{x}\|\leq k}{\ket{\vec{x}}}}\right\}
\frac{\ket{0}-\ket{1}}{\sqrt{2}}~.
\end{eqnarray}
Where $M_k$ is the appropriate normalisation factor calculated by the number
of $\vec{x}$ strings that have Hamming  weight less than or equal to $k$:
\begin{eqnarray} \label{eq:Mk}
M_k & = & \sum_{i=0}^{k}{\left({
\begin{array}{c}
N \\ i
\end{array}
}\right)}~.
\end{eqnarray}

Applying the above-described protocol $A_k$ (Equation~\ref{eq:Ak}) 
to this state yields (following Equation~\ref{eq:pkb}, and 
requiring $k$ oracle calls):
\begin{eqnarray} \label{eq:AkPsi}
A_k\ket{\Psi_k}\frac{\ket{0}-\ket{1}}{\sqrt{2}} & = & 
\frac{1}{\sqrt{M_k}}\left\{{
\sum_{\vec{x} \in \{0,1\}^N}^{\|\vec{x}\|\leq k}{
{(-1)}^{\inner{x}{\omega}}\ket{\vec{x}}}
}\right\}
\frac{\ket{0}-\ket{1}}{\sqrt{2}}~. 
\end{eqnarray}
Here we see how the phases of the state $A_k\ket{\Psi_k}$ 
contain a part of the desired information about 
$\omega_1\cdots\omega_N$ (like Equation~\ref{eq:Hinner} does 
for $y_1\cdots y_N$).

If we set $k$ to its
maximum $k=N$, then, applying an $N$-fold Hadamard to the first 
$N$ qubits of $A_k\ket{\Psi}$ would give us exactly the
state $\ket{\omega_1\cdots\omega_N}$. Whereas the minimum 
value $k=0$ leads to a state that does not reveal anything 
about $\vec{\omega}$.
For all the other possible values of $k$ between $0$ and $N$, 
there will be the situation that applying $H^{\otimes N}$ to the 
$\vec{x}$-register of $A_k\ket{\Psi_k}$ gives a state that is
close to $\ket{\omega_1\cdots\omega_N}$, but not exactly.
For a given $N$, this \emph{fidelity\/} (statistical correspondence) 
between the acquired state and $\vec{\omega}$ depends on $k$: as $k$
gets bigger, the fidelity increases.

\subsection{Analysis of the Algorithm}
The $N$ qubits that should give $\omega_1 \cdots \omega_N$ after 
the $H^{\otimes N}$ transformation, is described by 
(see Equation~\ref{eq:AkPsi}):
\begin{eqnarray} \label{eq:Psiprime}
\ket{\Psi_k'} & = & 
\frac{1}{\sqrt{M_k}}
\sum_{\vec{x} \in \{0,1\}^N}^{\|\vec{x}\|\leq k}{
{(-1)}^{\inner{x}{\omega}}\ket{\vec{x}}}~.
\end{eqnarray}
The probability that this state gives the correct string of $\omega$-bits
equals its fidelity with the perfect state $\ket{\Psi_N'}$:
\begin{eqnarray}
\textrm{Prob}(A_k \textrm{ outputs }\vec{\omega}) 
& = & {|\langle\Psi_k'|\Psi_N' \rangle|}^2 
\end{eqnarray}
The signs of the amplitudes of $\ket{\Psi_k'}$ and $\ket{\Psi_N'}$ 
will be the same for
all registers $\vec{x}$ with $\|\vec{x}\|\leq k$, whereas 
for the other strings with $\|\vec{x}\|>k$ the amplitudes of
$\ket{\Psi'_k}$ are zero.
The fidelity between the two states can therefore be calculated 
in a straightforward way, yielding for the correctness probability
(using Equation~\ref{eq:Mk}):
\begin{eqnarray} \label{eq:fidanalyze}
\textrm{Prob}(A_k \textrm{ outputs }\vec{\omega}) & = & 
\frac{M_k}{2^N} \nonumber \\
& = & 
\frac{1}{2^N}
\sum_{i=0}^{k}{
\left({
\begin{array}{c}
N \\ i
\end{array}
}\right)
}
\end{eqnarray}
this equality shows the reason why the algorithm also works for values of
$k$ around $n/2+\sqrt{n}$. for large $n$ the binomial distribution approaches
the Gaussian distribution. The requirement that the correctness probability 
has some value significantly greater than $1/2$, translates into the 
requirement that $k$ has to be bigger than the average $N/2$
by some multiple of the standard deviation $\sqrt{N}/2$ of the Hamming 
weights over the set of bit strings $\{0,1\}^N$.
Following this line of reasoning, it can be shown that
\begin{eqnarray}
\textrm{Prob}(A_{\lfloor N/2 +\sqrt{N}\rfloor}\textrm{ outputs }\vec{\omega})
& > & 0.95
\end{eqnarray}
for any value of $N$.

This proves that the following algorithm will give us the requested
$N$ oracle values $\omega_1 \cdots \omega_N$ with an error-rate
of less than $5\%$, using only $\lfloor N/2 + \sqrt{N}\rfloor$ queries
to the oracle.
\begin{enumerate}
\item{\textbf{Initial state preparation:} Prepare a register of $N+1$ 
qubits in the  state 
$\Psi_{\lfloor{N/2+\sqrt{N}}\rfloor}(\ket{0}-\ket{1})\sqrt{2}$ as 
in Equation~\ref{eq:Psi}.}
\item{\textbf{Oracle calls:} Apply the $A_k$ procedure of 
Equation~\ref{eq:Ak}, for $k=\lfloor N/2 +\sqrt{N}\rfloor$
oracle queries.}
\item{\textbf{Hadamard transformation:} Perform $N$ Hadamard transforms to
the first $N$ qubits on the register (the state
$\ket{\Psi_k'}$ in Equation~\ref{eq:Psiprime}).}
\item{\textbf{Final observation:} Observe the same first $N$ qubits in the
standard basis $\{\ket{0},\ket{1}\}$. The outcome of this observation
is our guess for the oracle description $\omega_1\cdots\omega_N$. 
This estimation of $\omega_1\cdots\omega_N$ will be correct \emph{for all\/} 
$N$ bits with a probability greater than $95\%$.}
\end{enumerate}
An expected error-rate of significantly less than $5\%$ can easily 
be obtained if we increase the threshold $k$ with a multiple of the 
`standard deviation' $\sqrt{N}/2$. The standard 
approximations of the binomial distribution by the Gaussian distribution 
shows that for big $N$ the error-probability goes to zero as the threshold
increases, according to the exponential relation:
\begin{eqnarray}
\textrm{Prob}_\mathrm{error}(k =
{N/2+\lambda\sqrt{N}}) & \approx &  
\frac{1}{2}-\frac{1}{2}\mathrm{Erf}(\sqrt{2}\lambda) \nonumber \\
& = & O\left({2^{-\lambda^2}}\right)~. 
\end{eqnarray} 
It is therefore that we can say that an arbitrary small error 
probability can be achieved with only $N/2+O(\sqrt{N})$ oracle calls.

\subsection{Comparison with Classical Algorithms}
Consider now a classical computer $B_k$ that is allowed to query the oracle 
$k$ times. This implies that after the procedure $N-k$ bits of 
$\vec{\omega}$ will be still unknown. Hence the probability of 
guessing the correct $N$-bit string $\omega_1\cdots\omega_N$
by any classical algorithm is:
\begin{eqnarray}
\textrm{Prob}(B_k\textrm{ outputs }\vec{\omega}) & \leq & 
\frac{1}{2^{N-k}}~.
\end{eqnarray}
This shows, as expected, that a classical probabilistic computer needs all 
$N$ oracle calls to obtain $\vec{\omega}$ with high probability.
The space complexity of the quantum and the classical algorithms
is in both cases linear in $N$.

\section{Approximate Interrogation}
In this section we ask ourself what happens if we want to know only
a certain \emph{fraction\/} of the $N$ unknown bits. In other words:
Given a threshold of $k$ oracle-queries, what is the maximum
expected number of correct bits $c$ that we can obtain via an 
`approximate interrogation' procedure if we assume $\vec{\omega}$ to
be totally random?

\subsection{Classical Approximate Interrogation}
In the classical setting the analysis is again straightforward.
If we query $k$ out of $N$ bits, then we know $k$ bits with
certainty and we have to randomly guess the other $N-k$ bits
of which we can expect 50\% to be correct.
The total number of correct bits will therefore be
\begin{eqnarray} \label{eq:clas_alg}
c_{k}^{\mathrm{clas}} & = & 
\frac{N}{2}+\frac{k}{2}~.
\end{eqnarray}
This shows a linear relation between $k$ and $c$.

\subsection{Quantum Approximate Interrogation}
The quantum procedure for approximate interrogation
will be the same algorithm as we used in the first part 
of the article, but with a different initial state $\Psi$. 
We now allow the amplitudes $\alpha_j$ of $\Psi$ to depend on the 
Hamming weight of its basis states $\vec{x}$. We therefore
write for the initial state:
\begin{eqnarray} \label{eq:Psi_approx}
\ket{\Psi_k^\alpha} & = & 
\sum_{j=0}^{k}{\alpha_j \cdot \frac{1}
{\sqrt{\left({\begin{array}{c}N \\ j\end{array}}\right)}}
\sum_{\vec{x}\in\{0,1\}^N}^{\|\vec{x}\|=j}
\ket{\vec{x}}}~,
\end{eqnarray}
with the normalisation restriction $\sum_j \alpha_j^2=1$.

After the preparation of this state $\Psi$, the algorithm is continued 
in the same way as described in Section~\ref{section:quantum_algorithm}.
The $N$ bits outcome of this protocol will correspond to a certain 
degree with the interrogated bit string $\omega_1\cdots\omega_N$,
where this degree depends on $k$ and the amplitudes $\alpha_j$.

In the appendix it is shown that the expected number of correct
bits for the quantum protocol will be
\begin{eqnarray} \label{eq:optimal}
c_{k}^{\mathrm{quant}} & = & 
\frac{N}{2} + \sum_{j=0}^{k-1}{\alpha_j\alpha_{j+1}\sqrt{j+1}\sqrt{N-j}}~.
\end{eqnarray}
For a given $k$ we can thus optimise the $\alpha$ values such that
$c_k$ will be as big as possible. 
Two examples of such optimisations will be given below,
both of them showing an improvement over the classical algorithm.

\subsection{Interrogation with One Quantum Query}
If we allow the quantum computer to ask only one query to the 
oracle, then Equation~\ref{eq:optimal} is maximised by choosing
$\alpha_0 = \alpha_1 = 1/\sqrt{2}$, thus giving for
the expected number of correct bits
\begin{eqnarray}
c_1^{\mathrm{quant}} & = & \frac{N}{2}+\frac{\sqrt{N}}{2}~.
\end{eqnarray}
When we compare this with Equation~\ref{eq:clas_alg}. we 
see that a classical algorithm would require $k=\sqrt{N}$ 
queries to match the power of a single quantum query. 

\subsection{Interrogation with Many Queries}
Let us assume that $N$ is big (such that $\sqrt{N}/N \approx 0$)
and that $k$ is a fraction of $N$ with $0\leq k/N \leq 1/2$.
We can then define the amplitudes $\alpha$ according to
\begin{eqnarray}
\alpha_j & = & \left\{\begin{array}{llrl}
0 & \mathrm{if~} & 
0 \leq j \leq & \!\!\!\!\!k-\sqrt{k} \\
\frac{1}{\sqrt[4]{k}} & \mathrm{if~} & 
k-\sqrt{k} < j \leq & \!\!\!\!\!k\\
\end{array}\right.
\end{eqnarray}
This gives us for the expected \emph{ratio\/} of correct bits
\begin{eqnarray}
\frac{c_{k/N}^{\mathrm{quant}}}{N} 
& = & 
\frac{1}{2} + \frac{1}{N\sqrt{k}}
\sum_{j=k-\sqrt{k}}^{k-1}
{\sqrt{j+1}\sqrt{N-j}} \nonumber \\
& \approx & 
\frac{1}{2}+\sqrt{\frac{k}{N}\left({1-\frac{k}{N}}\right)}~,
\end{eqnarray}
whereas for $1/2 < k/N \leq 1$ we use the $\alpha$ amplitudes
as if $k=N/2$ (with $c_{k/N} \approx N$).
 
\begin{figure}
\includegraphics{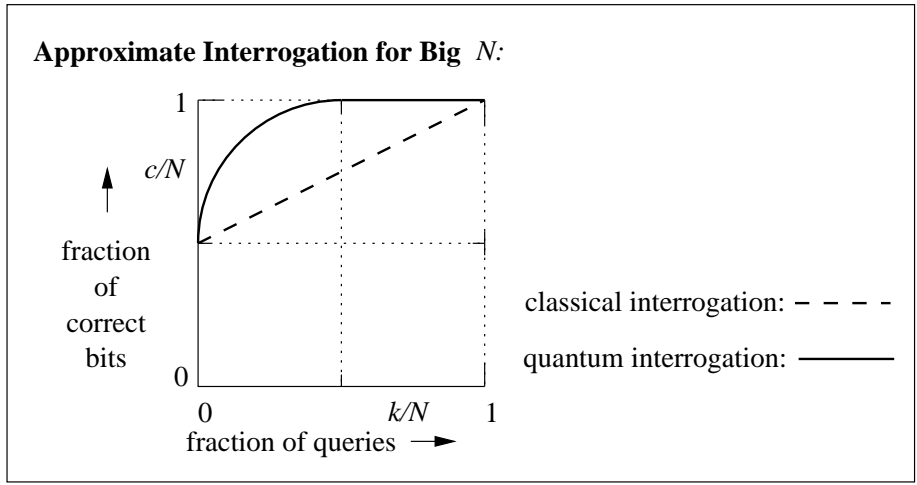}
\caption{Comparison of the interrogation effectiveness between classical and 
quantum computers.}
\label{figure:tradeoff}
\end{figure}

In the same setting, the classical fraction of correct bits will be
\begin{eqnarray}
\frac{c^{\mathrm{clas}}_{k/N}}{N} & = & 
\frac{1}{2}+\frac{k}{2N}~.
\end{eqnarray}
Again we see (Figure~\ref{figure:tradeoff}) that the quantum 
algorithm performs better than the classical one, especially 
for the small values of $k/N$. As an example: If we allow the
quantum protocol $N/10$ queries, then we can expect 80\% of
the bits to be correct. Any classical computer would need
$6N/10$ queries to obtain such a ratio. 

\section{Conclusions}
The model of quantum computation does not permit a general significant 
speed-up of the existing classical algorithms. Instead, we have to investigate
for each different kind of problem whether there is a possible 
gain by using quantum algorithms or not.

Here it has been shown that for every binary function $\omega$
with domain size $N$, we can obtain the full description of the function
with high probability while querying $\omega$ only 
$N/2 + \sqrt{N}$ times. A classical computer always requires
$N$ calls to determine $\omega_1\cdots\omega_N$ with the same kind
of success probability.

The lower bounds on \textsc{parity} (with bounded error) and
\textsc{or} (with no allowed error) for black-boxes\cite{Beals,Farhi}
show us that any quantum algorithm \emph{must\/} use at least $N/2$ calls
to obtain $\omega$ with bounded error, and that the full $N$ queries are 
necessary to determine the string without error, respectively.
The question that remains therefore, is if the $\sqrt{N}$-term in the 
query complexity $N/2+\sqrt{N}$ is necessary 
or can perhaps be reduced to the order of $\log N$, for example.

The term `approximate interrogation' was used for the scenario where
we are interested in obtaining a certain fraction of the $N$ unknown
bits. Again we could see how a quantum procedure outperforms 
the possible classical algorithms (Figure~\ref{figure:tradeoff}). 

For all results in this article we assumed $\omega$ to be a 
random oracle without any structure. Future research on quantum
computational complexity could investigate similar questions
for \emph{structured\/} oracles (the white-box model).
This might lead to results that widen the gap between classical and 
quantum computation even further than we did here.

\section*{Acknowledgements}
I would like to thank Harry Buhrman, Miklos Santha, 
Ronald de Wolf, Mike Mosca, and Artur Ekert
for useful conversations on this subject, and the latter three also 
for their critical proofreading of earlier versions of
this article.

This work was supported by the European TMR Research Network 
ERP-4061PL95-1412, Hewlett-Packard, and the Institute for Logic, 
Language, and Computation in Amsterdam.

\appendix
\section{Appendix: The Expected Number of Correct Bits
for the Quantum Algorithm}
In this appendix we will calculate how many bits we can expect to
be correct for the quantum interrogation procedure with the 
initial state $\Psi$ of Equation~\ref{eq:Psi_approx}.
We do this by assuming that the unknown bit string consists
of zeros only: $\vec{\omega} = 0\cdots 0$. 
The expected number of correct bits for the algorithm 
equals therefore the expected number of zeros of the observed output 
string $\vec{y}$.
Because we can make the assumption $\vec{\omega}=\vec{0}$
 without loss of generality, 
we can then afterwards conclude that this
number will the expected number of correct bits for any 
$\vec{\omega}$.

The inner-product between $\vec{x}$ and $\vec{\omega}$ 
will be zero for every $\vec{x}$,
hence applying 
$A_k$ to $\Psi$ will not change the initial state:
\begin{eqnarray}
A_k \ket{\Psi_k^\alpha} & = & 
\sum_{j=0}^{k}{\alpha_j \cdot \frac{1}
{\sqrt{\left({\begin{array}{c}N \\ j\end{array}}\right)}}
\sum_{\vec{x}\in\{0,1\}^N}^{\|\vec{x}\|=j}
\ket{\vec{x}}}~.
\end{eqnarray}
After this $A_k$, we perform the $N$ Hadamard transforms on all 
$N$ qubits, yielding a new state:
\begin{eqnarray}
{H^{\otimes N}A_k\ket{\Psi_k^\alpha}} 
&= & \sum_{j=0}^{k}{\frac{\alpha_j}
{\sqrt{\left({\begin{array}{c}N \\ j\end{array}}\right)}}
\sum_{\vec{x}\in\{0,1\}^N}^{\|\vec{x}\|=j}
H^{\otimes N}\ket{\vec{x}}} \nonumber \\
& = & 
\frac{1}{\sqrt{2^N}}
\sum_{\vec{y}\in\{0,1\}^N}
\sum_{j=0}^{k}{\frac{\alpha_j}
{\sqrt{\left({\begin{array}{c}N \\ j\end{array}}\right)}}
\sum_{\vec{x}\in\{0,1\}^N}^{\|\vec{x}\|=j}
{(-1)}^{\inner{y}{x}}\ket{\vec{y}}} 
\end{eqnarray}
Because the above state is invariant under permutation,
the probability of observing a certain string $\vec{y}$ depends
only on its Hamming weight $\|\vec{y}\|$. This, in combination 
with some other known equalities\cite{Gradshteyn} and mathematical 
techniques, enables us to express the expected number of zeros by
\begin{eqnarray}
\textrm{\#zeros}({
{H^{\otimes N}A_k\ket{\Psi_k^\alpha}}
}) & = & 
\sum_{t=0}^{N}
{t\cdot \left({\begin{array}{c}N \\ t\end{array}}\right) 
{|\langle0^t1^{N-t}|
{H^{\otimes N}A_k\ket{\Psi_k^\alpha}}|}^2} \nonumber \\
& = & 
\frac{N}{2} + \sum_{j=0}^{k-1}{\alpha_j\alpha_{j+1}\sqrt{j+1}\sqrt{N-j}}~.
\end{eqnarray}
We can therefore conclude that the expected number $c_k$ of correctly
guessed bits for the quantum protocol will be
(for given $k$ and $\alpha_j$):
\begin{eqnarray} 
c_k^{\mathrm{quant}} & = & 
\frac{N}{2} + \sum_{j=0}^{k-1}{\alpha_j\alpha_{j+1}\sqrt{j+1}\sqrt{N-j}}~.
\end{eqnarray}
\end{document}